\documentclass[a4paper,11pt]{iopart}
\usepackage{graphicx}
\usepackage{cite}
\usepackage{color}
\usepackage{amsfonts}
\usepackage{srcltx}
\begin{document}
\title{Identifying universality classes of absorbing phase transitions by block renormalization}
\author{Urna Basu$^1$ and Haye Hinrichsen$^2$}
\address{$^1$Theoretical Condensed Matter Physics Division, Saha Institute of Nuclear Physics, 
1/AF Bidhan Nagar, Kolkata 700064, India}
\address{$^2$ Universit\"at W\"urzburg, Fakult\"at f\"ur Physik und Astronomie, Am Hubland, \\ 97074 W\"urzburg, Germany}

\ead{urna.basu@saha.ac.in}

\begin{abstract}
We propose a renormalization scheme that can be used as a reliable method to identify universality classes of absorbing phase transitions. Following the spirit of Wilson's block-spin renormalization group, the lattice is divided into blocks, assigning to them an effective state by a suitable Boolean function of the interior degrees of freedom. The effective states of adjacent blocks form certain patterns which are shown to occur with universal probability ratios if the underlying process is critical. Measuring these probability ratios in the limit of large block sizes one obtains a set of universal numbers as an individual fingerprint for each universality class.
\end{abstract}

\def\smallcaption#1{\caption{\footnotesize #1}}
\def\headline#1{\noindent\textbf{\small#1}\\[3mm]}
\def\P#1{P_{\tt#1}}
\def\S#1{S_{\tt#1}}
\def\n{\nonumber}

\pagestyle{plain}

\section{Introduction}

Scale invariance is known to be an important cornerstone for our understanding of critical phenomena and provides the basis of renormalization group theory. On a lattice a scale transformation can be carried out by coarse-graining the elementary degrees of freedom. One of the best known examples is Wilson's block-spin renormalization of the Ising model, where several spins are grouped into blocks~\cite{Fisher,Wilson,Kadanoff}. If the Hamiltonian in terms of these block variables has the same form as the original one (possibly with different parameters), this procedure maps the system onto itself, forming a group of renormalization transformations. Critical phenomena are typically associated with fixed points under renormalization group transformations. 

Non-equilibrium systems differ from their equilibrium counterparts in so far as they involve time as an additional dimension on equal footing with the spatial degrees of freedom. Nevertheless the situation is similar, i.e. by appropriately coarse-graining the elementary degrees of freedom we can identify critical phenomena as fixed points under renormalization group transformations. In the past decades various renormalization techniques for nonequilibrium systems  have been developed, including field-theoretic renormalization methods~\cite{Cardy}, density matrix renormalization~\cite{Peschel}, and continuous non-perturbative renormalization techniques~\cite{Canet}, to name only a few. 

In the present paper we consider a discrete renormalization scheme in the spirit of Wilson's block-spin method that is suitable for systems with continuous nonequilibrium phase transitions into absorbing states~\cite{Review,Odor,Luebeck,Henkel}. The most important example of such systems is directed percolation (DP)~\cite{Kinzel}, a  universality class of absorbing phase transitions which plays a paradigmatic role similar to the Ising class in equilibrium statistical mechanics. Other examples include the class of compact directed percolation, the parity-conserving universality class, voter-type phase transitions, the pair-contact process with diffusion, and the Manna class (for a review see e.g.~\cite{Henkel}). Universal behavior is also observed at tricritical points~\cite{Lubeck,Grassberger} and in models with algebraic long-range interactions~\cite{Levy}. Recently absorbing phase transitions attracted increasing attention since the critical behavior of directed percolation could be reproduced experimentally for the first time in turbulent liquid crystals~\cite{Takeuchi}.

Before going into detail let us briefly sketch how the  proposed method for renormalization works. Restricting for simplicity to the 1+1-dimensional case, it involves the following steps:
\begin{itemize}
\item[-] Consider a sufficiently large system on a chain with $L$ sites and periodic boundary conditions starting with a homogeneously active state \textit{at criticality}.\\[-3mm]
\item[-] Read out the configuration of active sites at time $t$ and divide the chain into blocks of equal size~$b$. A block containing at least one active site is called active ({\tt 1}) and otherwise inactive ({\tt 0}), giving a string of $L/b$ bits.\\[-3mm]
\item[-] Scan the bit string cyclically in order to estimate the probability $\P c(b,t)$ to find $n$~randomly selected adjacent blocks with the bit pattern $\tt c$. For example, three adjacent blocks can have eight different patterns \ ${\tt c}\in \{000,001,010,\ldots,111\}$. 
\end{itemize}
This procedure is carried out at different times $t$ and for different block sizes $b$, averaging over many runs to reduce statistical fluctuations. Note that the underlying process is not affected, all what is needed is to read out the configurations in regular time intervals and to process this data in a separate procedure in order to estimate $\P c(b,t)$.

The probabilities $\P c(b,t)$ will depend on both the block size $b$ and the actual time~$t$. However, as will be shown below, the \textit{ratios} of these probabilities saturate and can be extrapolated to $t \to\infty$. These block-size-dependent extrapolated probability ratios in turn converge to well-defined \textit{universal} values in the limit $b\to\infty$, providing a whole set of universal numbers which can be viewed as a fingerprint of the underlying universality class. With these numbers denoted as $\S c$, which are as robust as critical exponents, it is possible to identify a universality class in a reliable way.

\section{Block renormalization scheme for systems with absorbing states}\label{sec:blockrenorm}

\headline{Symmetry as the guiding principle}
%
The purpose of any block renormalization scheme is to organize the elementary degrees of freedom (the lattice sites) into equal groups and to replace the individual states inside each group by a single effective state of the same kind. This coarse-graining has to be done in such a way that the symmetries of the model are preserved. For example, in Wilson's block renormalization scheme for the one-dimensional Ising model the spins are grouped in pairs. One of the two spins is taken as the effective spin of the block while the other spin is integrated out and absorbed into the effective interaction of two neighboring blocks. Obviously, this procedure preserves the global $Z_2$-symmetry of the Ising model.

Devising a similar method for models with absorbing state transitions, the coarse-graining procedure should respect the most salient feature of such models, namely, the existence of an absorbing state. An absorbing state is a configuration that can be reached but not be left and therefore plays the role of a dynamical trap. Consequently any renormalization procedure should map a locally absorbing configuration of lattice sites onto an absorbing effective state of the block and conversely every non-absorbing configuration onto an active effective state. In other words, the renormalization procedure has to preserve the property of `being absorbing'.\\

\headline{Two-state models: Renormalization by a logical OR}
%
Many models with absorbing phase transitions are two-state models, where each lattice site $i$ carries a binary bit~$s_i\in\{${\tt 0},{\tt 1}$\}$, denoting inactive and active sites, respectively. The renormalization procedure outlined above groups these bits into equal-sized blocks, assigning to each block an effective state in form of a single bit which is again either {\tt 0} or {\tt 1}. Therefore, the desired map has to be a Boolean function. 

Since the renormalization procedure has to preserve the absorbing state, a locally absorbing configuration {\tt 00...0} has to be mapped to {\tt 0}. Conversely, any non-absorbing configuration, where the block contains at least one active site, has to be mapped to {\tt 1}. Obviously, the only Boolean function with this property is a logical OR. Thus we can conclude that any block renormalization scheme for two-state systems with  a single absorbing state should be based on a logical OR of the individual degrees of freedom.

\begin{figure}
\centering\includegraphics[width=130mm]{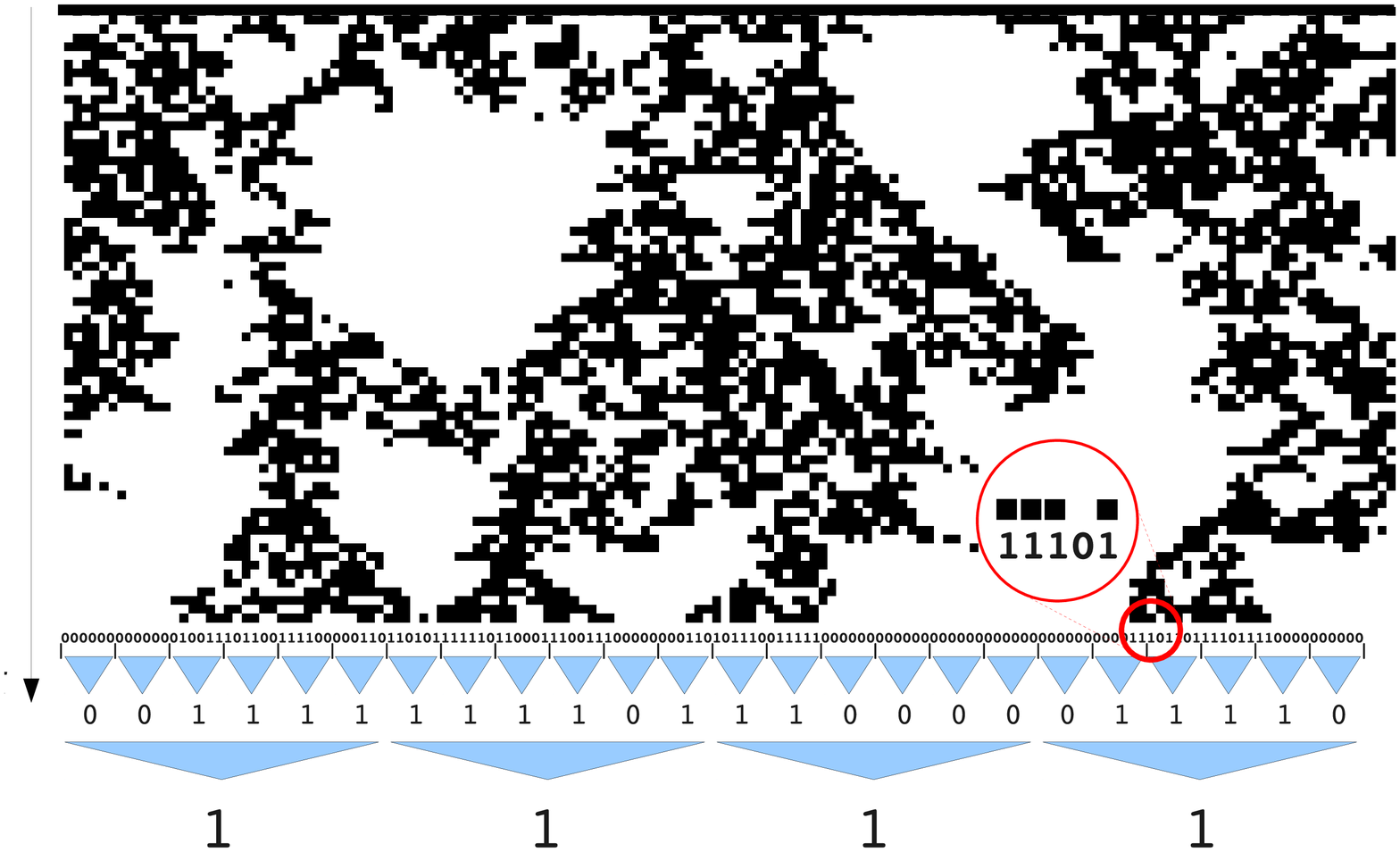}
\smallcaption{Block renormalization of two-state models with absorbing states. The figure shows a contact process at criticality starting with a fully occupied lattice. The upper part shows the temporal evolution, where each row of pixels represents the configuration after 10 Monte Carlo sweeps. At a certain point we read out the spatial configuration of actives sites which can be considered as a string of bits. The figure shows two subsequent renormalization steps symbolized by blue triangles, each dividing the chain into groups of six bits and mapping them onto a single bit by a logical OR. Obviously these two steps could be replaced by a single one, grouping 36 bits together, demonstrating that these operations form a commutative and associative group.}
\label{fig:rg}
\end{figure}

The coarse-graining procedure by a logical OR can be implemented in different ways. In the following we will use it to renormalize spatial configuration (snapshots) of the system at a certain instance of time, as sketched in Fig.~\ref{fig:rg}. After renormalization one obtains coarse-grained bit patterns of adjacent block variables. The aim is to analyze their probability distribution, to study possible universal properties, and to use the results in order to identify the underlying universality class of the process.\\

\headline{Scaling properties under coarse graining}
%
The critical properties of absorbing phase transitions are well described by a phenomenological scaling theory~\cite{Henkel}. Starting point is the postulate that such systems are invariant under anisotropic scale transformations
\begin{equation}
\label{rescaling}
x \to \Lambda x, \qquad t \to \Lambda^z t, \qquad \rho \to \Lambda^{-\beta/\nu_\perp} \rho, \qquad \Delta \to \Lambda^{- 1/\nu_\perp} \Delta \,,
\end{equation}
where $\Lambda>0$ is a spatial scale factor, $\rho$ denotes the coarse-grained density of active sites, and $\Delta=\lambda-\lambda_c$ denotes the distance from criticality in the control parameter $\lambda$. Apart from few exceptions~\cite{Rittenberg,Levy}, time and space scale differently, expressed by a dynamical exponent $z=\nu_\parallel/\nu_\perp>1$. Thus the scaling behavior of such a system is determined by three independent critical exponents $\beta,\nu_\perp,\nu_\parallel$ from which all other bulk exponents can be derived by simple scaling relations\footnote{In principle there could be a fourth independent exponent $\beta'$ which is associated with the conjugate order parameter. However, in most cases a time reversal symmetry implies $\beta=\beta'$, see e.g. Ref.~\cite{Henkel}.}. 

By grouping $b$ lattice sites into blocks the lattice spacing $a$ increases by $a \to ab$. This means that the spatial coordinate $x$ in Eq.~(\ref{rescaling}), which measures the distance in units of the lattice spacing, decreases by $x \to x/b$. Therefore, the scale factor of each renormalization step is $\Lambda=1/b$ so that the quantities discussed above are expected to scale as
\begin{equation}
\label{rescaling2}
x \to x/b, \qquad t \to b^{-z} t, \qquad \rho \to b^{\beta/\nu_\perp} \rho, \qquad \Delta \to b^{1/\nu_\perp} \Delta \,.
\end{equation}
Note that the density of active blocks increases under coarse-graining, reflecting the fact that a logical OR can only increase the density of active blocks, as illustrated in Fig.~\ref{fig:rg}. Moreover, the last equation $\Delta \to b^{1/\nu_\perp} \Delta$ tells us that the distance from criticality effectively increases under renormalization, driving the system away from criticality. \\

\headline{Scaling properties of the probability distribution $\P c(b,t)$}
%
Suppose that we have read out the configuration of the system at time $t$ and arranged it into blocks of size $b$ as shown in Fig.~\ref{fig:rg}. Scanning over the resulting bit pattern (and averaging over several runs) we can now estimate the probability $\P c(b,t)$ to find $n$ randomly selected adjacent blocks in the configuration ${\tt c}=\{c_1c_2\ldots  c_n\}$. By definition, these probabilities are normalized by $\sum_{\tt c} \P c(b,t)=1$. 

How do these probabilities decay asymptotically as functions of time? To answer this question we first note that $\P {1}(1,t)$ is just the density of active sites on the original lattice, hence this quantity is known to decay as
\begin{equation}
\P {1}(1,t) = \rho(t) \sim t^{-\alpha}\,,
\end{equation}
where $\alpha=\beta/\nu_\parallel$. If the renormalization scheme outlined above works properly, the same should hold for all single-bit quantities after coarse-graining, i.e. we expect them to decay asymptotically as
\begin{equation}
\P {1}(b,t) \sim t^{-\alpha}
\end{equation}
for arbitrary block sizes $b$, differing only by a $b$-dependent proportionality factor in front of the power law. This observation can be confirmed easily by numerical simulations.

Let us now turn to the two-bit probabilities. On the one hand, $\P {00},\P {01},\P {10},$ and $\P {11}$ are restricted by the normalization condition and the left-right symmetry
\begin{eqnarray}
&& \P {00}(b,t)+\P {01}(b,t)+\P {10}(b,t)+\P {11}(b,t)=1\,, \\
&& \P {01}(b,t)=\P {10}(b,t)\,.
\end{eqnarray}
On the other hand, the renormalization mapping by a logical OR can be expressed as
\begin{equation}
\P {01} (b,t) + \P {10} (b,t) + \P {11} (b,t) = \P {1} (2b,t) \sim t^{-\alpha}\,.
\label{eq:norm}
\end{equation}
Finally, tracing out the second bit one obtains the relation
\begin{equation}
\P {10} (b,t) + \P {11} (b,t) = \P {1} (b,t) \sim t^{-\alpha}
\end{equation}
with the same power law as in (\ref{eq:norm}) but with a smaller proportionality factor. These four equations imply that the two-bit probabilities with non-vanishing bit patterns (i.e. $\P {01} (b,t)$, $\P {10} (b,t)$, and $\P {11} (b,t)$) decay asymptotically in the same way as the density of active sites $\rho(t)$, but with different  prefactors in front of the power law.

\begin{figure}
\centering\includegraphics[width=155mm]{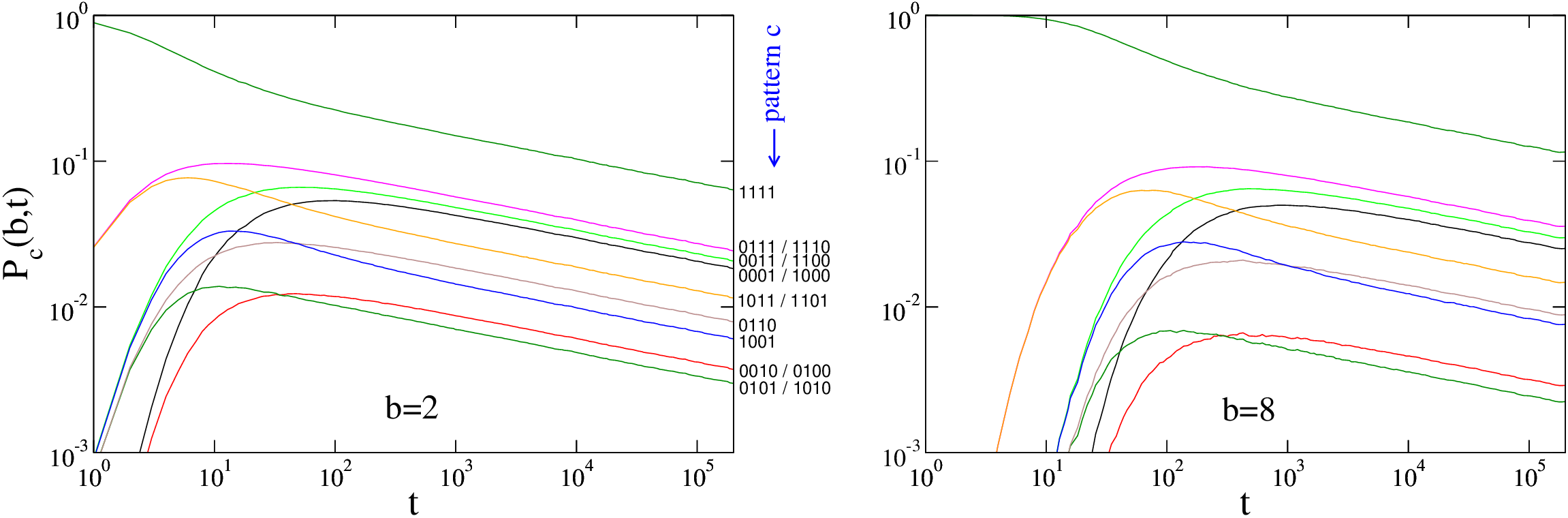}
\smallcaption{Temporal decay of the four-bit probabilities $\P c(b,t)$ for two different block sizes $b=2$ and $b=8$ in the contact process. Note that all quantities decay asymptotically with the same power law.}
\label{fig:decaydemo}
\end{figure}

Similarly all $n$-bit probabilities for arbitrary block sizes $b$ are found to decay asymptotically in the same way as the density of active sites, i.e.,
\begin{equation}
\P c(b,t) \sim  t^{-\alpha}  \quad \mbox{ if } {\tt c}\neq{\tt 000\ldots 0}\,.
\end{equation}
This is demonstrated in Fig.~\ref{fig:decaydemo} in the case of DP, where we plotted the decay of the four-bit probabilities for two different block sizes. Note that this common type of algebraic decay sets in only when the correlation length of the process exceeds the size of the block sequence, i.e. $\xi_\perp \gg nb$. This requires a simulation time $t \gg (nb)^z$, where $z=\nu_\parallel/\nu_\perp$ is the dynamical exponent of the process under consideration.\\

\headline{Time-independent quotients in the limit $t \to \infty$}
%
Since all probabilities $\P c(b,t)$ with  ${\tt c}\neq{\tt 00\ldots 0}$ decay asymptotically in the same way, any quotient of these quantities will saturate at some constant value in the limit $t\to\infty$. As a convenient choice we decided to study the ratios
\begin{equation}
\S c(b,t) \;:=\; \frac{\P c(t)}{\sum_{c'\neq{\tt 00\ldots0}} P_{c'}(t)} \;=\; \frac{\P c(t)}{1-\P{00\ldots0}(t)} \qquad ({\tt c}\neq{\tt 00\ldots 0})\,,
\end{equation}
where we used the normalization condition in the denominator. These quantities can be interpreted as the \textit{conditional} probability to find $n$ subsequent blocks of size~$b$ in the bit pattern ${\tt c}$, given that at least one of the blocks is active. By definition, they are normalized by
\begin{equation}
\sum_{c \;\neq\; {\tt 00\ldots 0}}  \S c(b,t)=1\,.
\end{equation}
As discussed before, the conditional probabilities tend to a constant
\begin{equation}
\label{eq:timelimit}
\S c(b) \;:=\; \lim_{t \to \infty}\S c(b,t)\,
\end{equation}
in the limit $t \to\infty$. In this way we have defined stationary block probabilities in a non-stationary process. \\

\begin{figure}[t]
\centering\includegraphics[width=110mm]{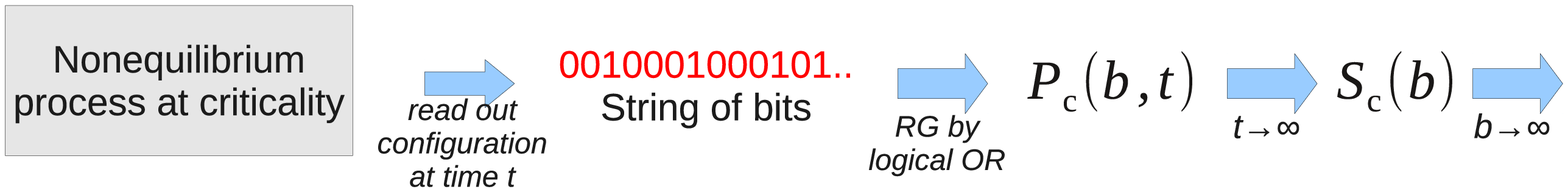}
\vspace*{-1cm}
\smallcaption{Main steps of the method proposed in the present work.
\label{fig:mainsteps}}
\end{figure}

\headline{Universality of the conditional probabilities in the limit $b \to \infty$}
%
Since the conditional probabilities $\S c(b)$ can be viewed as some kind of short-range correlation functions, we expect them to be non-universal, reflecting the microscopic details of the dynamics. However, the main message of this paper is the observation that the block renormalization procedure introduced above drives the conditional probabilities towards certain universal values which depend only on the large-scale critical behavior of the process under consideration. This means that the conditional probabilities converge to well-defined universal values 
\begin{equation}
\label{eq:limitblocksize}
\S c \;=\: \lim_{b \to \infty} \S c(b)
\end{equation}
in the limit of large block sizes. Note that this limit has to be carried out \textit{after} taking $t \to \infty$ since the two limits (\ref{eq:timelimit}) and (\ref{eq:limitblocksize}) do not commute. 

As will be demonstrated in the following section, it is possible to carry out the two-fold non-commuting limit in numerical simulations which allows one to estimate the numbers $\S c$, following the sequence of steps shown in Fig.~\ref{fig:mainsteps}. As mentioned before, these numbers can be viewed as a fingerprint of the underlying universality class. Since our method can be applied to any models with absorbing states, we can use these numbers to verify whether a given process belongs to a certain universality class or not.\\

\headline{Exact scaling relations for the two-bit probabilities}
%
As will be shown now, the universal two-bit probabilities $\S {01},\S {10},\S {11}$ are special in so far as they can be expressed exactly in terms of the critical exponents. To see this recall that $\P 1(1)$ is the density of active sites, while $\P 1(b)$ is the renormalized density of active sites, defined as the probability to find at least one active site in a block of size $b$. According to Eq.~(\ref{rescaling2}) this quantity is expected to scale as
\begin{equation}
\P 1(b) \sim \P 1(1) \, b^{\beta/\nu_\perp}
\end{equation}
as $b \to \infty$. This asymptotic power law implies the identity
\begin{equation}
\lim_{b \to \infty} \frac{\P 1(b)}{\P 1(2b)} \;=\; 2^{-\beta/\nu_\perp}\,.
\end{equation}
Using the exact relations $\P 1(2b) = 1- \P 0(2b) = 1- \P  {00}(b)$ 
and $\P 1(b) = \P {01}(b)+ \P {11}(b)$ this identity can be rewritten as
\begin{equation}
\lim_{b \to \infty} \frac{\P {01}(b)+ \P {11}(b)}{1- \P  {00}(b)} \;=\; \S {01} + \S {11} \;=\; 2^{-\beta/\nu_\perp}\,.
\end{equation}
Together with the normalization condition and spatial reflection symmetry we obtain a system of three equations
\begin{eqnarray}
\S {01} + \S {11}  = 2^{-\beta/\nu_\perp} \\
\S {01} + \S {10} + \S {11} = 1 \\
\S {01} = \S {10} 
\end{eqnarray}
with the exact solution
\begin{eqnarray}
\label{eq:exact1}
\S {01} = \S {10} = 1-2^{-\beta/\nu_\perp}  \\
\label{eq:exact2}
\S {11} = -1+2^{1-\beta/\nu_\perp} \,. 
\end{eqnarray}

If the critical exponents are known, this allows us to predict the two-bit quantities and to verify the accuracy of the method in a numerical simulation.
In the remaining part of this article we use Monte Carlo simulations to do the same for two well known universality classes of absorbing phase transitions, namely directed percolation and parity conserving class. 

Note that the conditional probabilities with three or more bits cannot be calculated in this way since the scaling laws do not provide enough equations. This is good news since these probabilities provide additional information in form of universal numbers which are independent of the usual critical exponents. From the numerical simulations we also estimate these universal numbers for the above mentioned classes.

\section{Directed percolation}

\headline{Verification of the two-bit probabilities}
%
For directed percolation in one dimension the best known estimates of the critical exponents are~\cite{Jensen}
\begin{equation} 
\beta = 0.276486(8)\,,\quad 
\nu_\perp = 1.096854(4)\,,\quad
\nu_\parallel = 1.7733847(6).
\end{equation}
\begin{figure}
\centering\includegraphics[width=150mm]{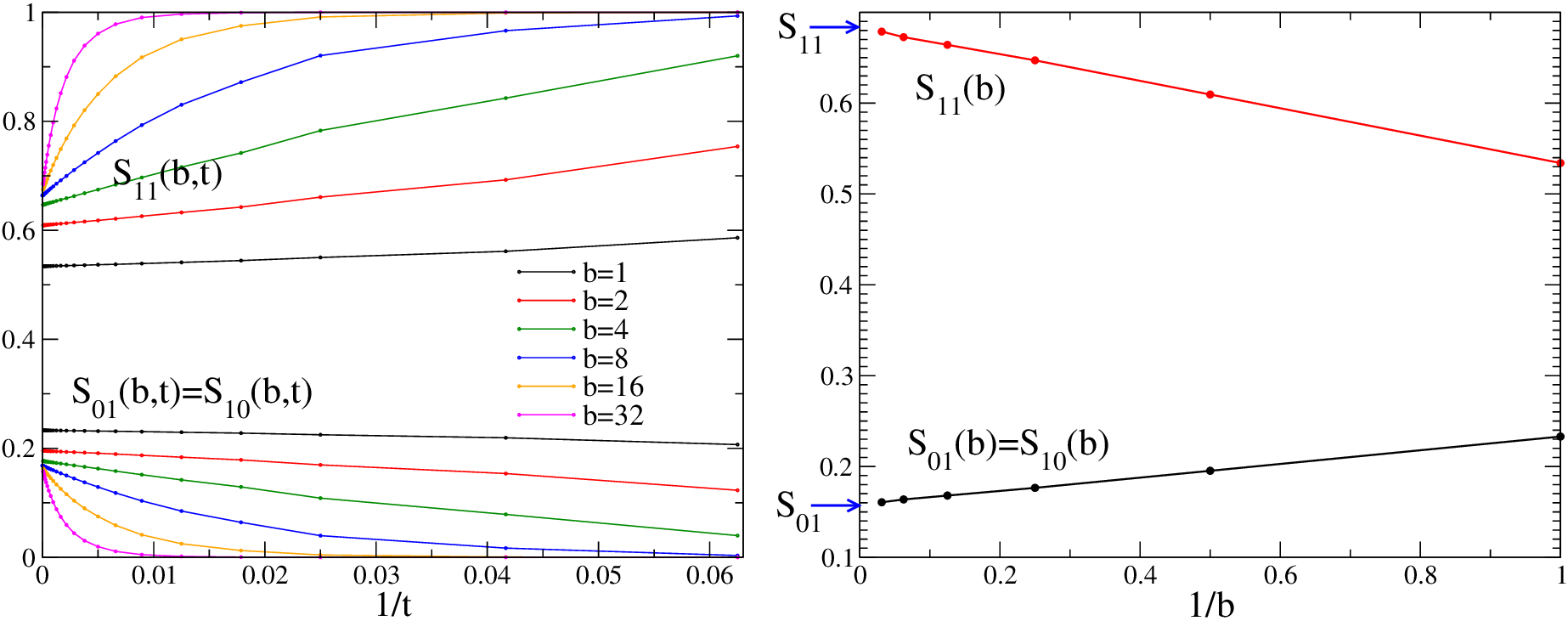}
\smallcaption{Two-bit probabilities for directed percolation. Left: Estimation of the conditional two-bit probabilities $\S {01}(b)$ and $\S {11}(b)$ in the limit $t \to \infty$  according to~(\ref{eq:timelimit}). Right: Extrapolation of the universal values $\S {01}$ and $\S {11}$ for $b \to \infty$ according to~(\ref{eq:limitblocksize}). As can be seen, the values converge to the predicted values in Eq.~(\ref{eq:exactnumerical}) which are indicated by the arrows.}
\label{fig:2bit}
\end{figure}
Inserting these estimates into Eq.~(\ref{eq:exact1}) and~(\ref{eq:exact2}) we can compute the conditional two-bit probabilities
\begin{equation}
\label{eq:exactnumerical}
\S {01} = \S {10} \simeq 0.160310(5) \,, \qquad
\S {11} \simeq 0.67938(1)\,. 
\end{equation}
Let us now demonstrate how these values can be obtained numerically by performing the two non-commuting limits $t\to\infty$ and $b\to\infty$. To this end we simulate an ordinary bond DP process starting with a fully occupied lattice over $10^4$ Monte Carlo sweeps. In certain intervals we read out the actual configuration and apply the suggested renormalization scheme, reducing $b$ bits to a single bit by a logical OR operation. Scanning over the resulting bit string and averaging over many runs, we estimate the conditional probabilities $\S c(b,t)$ as a function of time for the block sizes $b=1,2,4,8,16,$ and $32$. 

As shown in the left panel of Fig.~\ref{fig:2bit}, these probabilities tend to certain asymptotic values $\S c(b)$ as $t$ is increased. It turns out that the difference $\S c(b,t)-\S c(b)$ decays asymptotically as $1/t$, - a numerical observation for which we do not yet have an explanation. Hence, by plotting $\S c(b,t)$ versus $1/t$ we can easily extrapolate the curves to the limit $t \to \infty$. 

The extrapolated values for $\S c(b)$ are then plotted in the right panel of~Fig.~\ref{fig:2bit}. In order to perform the second limit $b \to \infty$ it is again useful to plot these values versus $1/b$ and to estimate the limit $\S c=\lim_{b \to \infty}\S c(b)$ by linear extrapolation. As can be seen, this procedure accurately reproduces the predicted values in Eq. (\ref{eq:exactnumerical}) which are indicated by arrows in the figure.  \\

\begin{figure}
\centering\includegraphics[width=110mm]{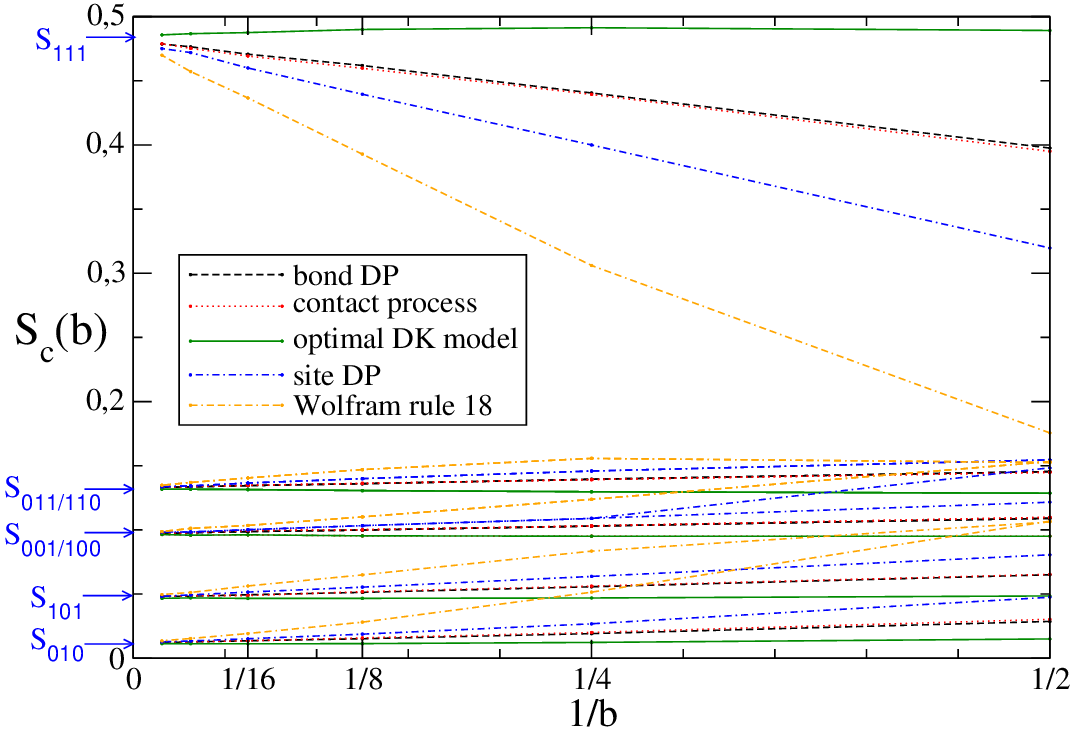}
\smallcaption{Confirmation of universality. The figure shows the 3-bit probabilities $\S c(b)$ with $b=2,4,8,16,32,64$ for various well-known lattice models of DP, namely directed bond percolation, the contact process, directed site percolation and Wolframs W18 stochastic cellular automaton. In addition, the figure shows data for the Domany-Kinzel model with optimized parameters (see text). As can be seen, all estimates tend to the same universal values as $b \to  \infty$. }
\label{fig:compare}
\end{figure}

\headline{Numerical confirmation of universality}
%
In order to demonstrate the universal character of the extrapolated conditional probabilities $\S c$ in the case of bit patterns with more than two bits, we measured the 3-bit quantities for various frequently used models in the directed percolation class, namely the contact process with random sequential updates, bond DP, site DP and Wolframs rule W18 (see~\cite{Henkel} and references therein). As shown in Fig.~\ref{fig:compare}, these models produce different values for $\S c(b)$ but they converge towards the same value as the block size is taken to infinity, confirming the universality hypothesis. Surprisingly the contact process and bond percolation produce almost identical results, indicating that they are probably characterized by very similar non-universal short-range correlations.

Moreover, we studied the Domany-Kinzel (DK) model~\cite{Domany} along its critical line parametrized by $p_1$ and $p_2$, searching for a point where the deviation for small block sizes is minimal (see Fig.~\ref{fig:compare}). Without going into detail we would like to report that this optimal point is located at $p_1=0.58640(2)$ and $p_2=0.95923(2)$. We expect that this transition point of the critical line is characterized by minimal corrections to scaling.

\section{Parity-conserving universality class}

Parity conserving (PC) models belong to a universality 
class of absorbing state transitions different from DP. The defining feature of this class of systems is 
that the dynamics conserves the number of particles modulo $2$
which is referred to as the {\em parity} of the system.
There are several variants of parity conserving models which are all characterized by the critical exponents~\cite{Henkel}
\begin{equation}
\beta = 0.92(2),\quad \nu_\perp = 1.83(3), \quad\nu_\parallel = 3.22(6). \label{eq:pc}
\end{equation}

To apply the block renormalization procedure for the PC class, one needs to make sure that
the renormalization scheme also respects the additional microscopic symmetry present in the system, {\it i.e.} the 
conservation of parity. This can be ensured if the reduction of a block of length $b$ to a single site keeps the parity of that block unchanged. 
This is not possible with a two-state model, where a bit obtained by reducing a block does not carry
any information about the parity. Instead one needs three states, namely, the inactive state and
two active states indicating whether a block contains an even or odd number of particles.

In this spirit we devise a parity-conserving three state model with an absorbing phase transition.
As will be discussed below, this model renders itself easily to the scheme 
of block renormalization without hampering the parity of the original configuration.

\begin{figure}
\centering\includegraphics[width=150mm]{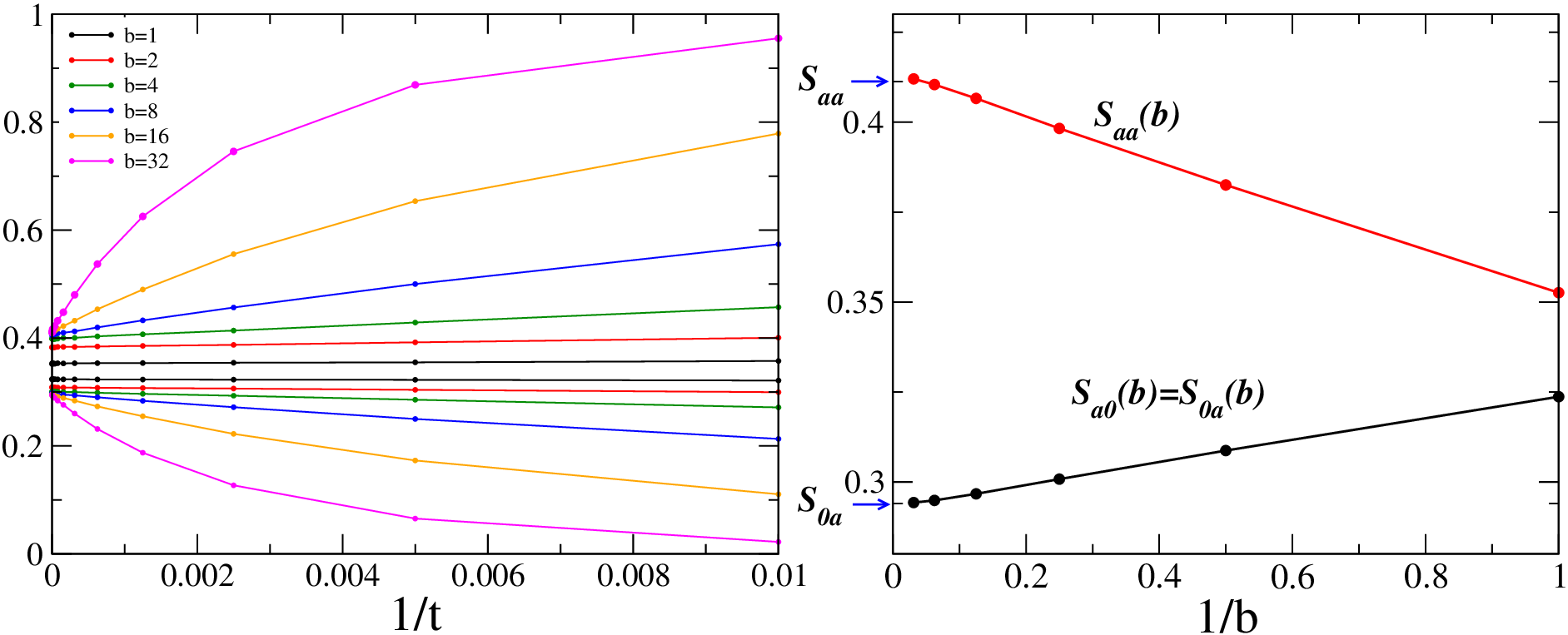}
\smallcaption{Two-block probabilities for the parity-conserving class. Left panel: $\S c (b,t)$ as a function of $1/t$ for various block sizes $b$ measured in the PC model defined in (\ref{eq:pcrules}). Right panel: Extrapolated values $\S c (b)=\lim_{t\to\infty} \S c(b,t)$ plotted against $1/b$. 
\label{fig:pc_2bit}}
\end{figure}

The model is defined on a one-dimensional periodic lattice, where each lattice site~$i$ is either empty 
or occupied by an odd or even number of particles, corresponding to
the state variables $s_i=0,1$ and $2$, respectively. The system evolves in time according to the following dynamical rules, 

\begin{eqnarray}
\left. 
\begin{array}{ccc}
 &~ 10 \to 21 &~ 20 \to 11 \cr 
01 \to 12 &~ 11 \to 22 &~ 21 \to 12 \cr 
02 \to 11 &~ 12 \to 21 &~ 22 \to 11 
\end{array}\right\} {\rm~with ~ rate~} p, \cr
\hspace{45mm} 2\to 0 \hspace{7mm} {\rm ~with ~ rate~} 1-p. \label{eq:pcrules}
\end{eqnarray}

\noindent This dynamics either adds or removes a pair particles, 
keeping the parity of the system conserved. The only absorbing configuration is $000...$ since any 
occupied site is considered as active. The order parameter $\rho$ is the density of active sites, not the 
density of particles in the system. This system shows an absorbing state transition belonging to the PC class defined by the critical 
exponents (\ref{eq:pc}) at the critical point 
\begin{equation}
 p_c=0.4017(1)\,.
\label{eq:pcpc}
\end{equation}
The parity-conserving variant of the block renormalization scheme 
can be applied to this model as follows. A block of length $b$ 
is replaced by a single site with the state 
\begin{quote}\begin{tabular}{ll}
0 & if the block is empty,\\
1 & if the total parity with the block is odd,\\
2 & if the total parity with the block is even.
\end{tabular}\end{quote}
This choice makes sure that the parity of the renormalized configuration 
is same as that of the original one. Of course, as in the previous case, 
the absorbing configuration of the original lattice still remains absorbing 
under this renormalization scheme.

\begin{figure}
\centering\includegraphics[width=100mm]{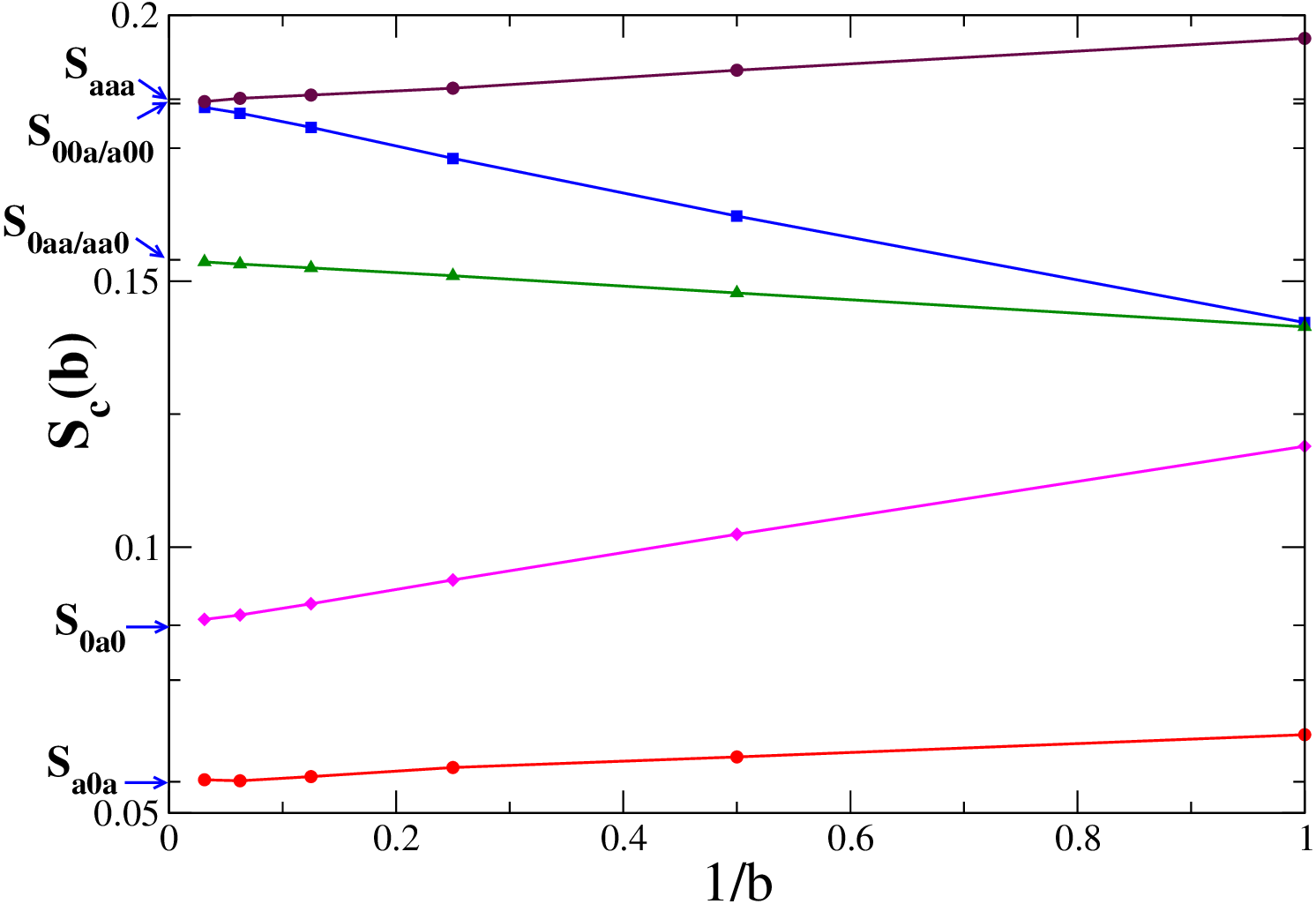}
\smallcaption{Extrapolated three-block probabilities $\S c(b)$ plotted against $1/b$. Extrapolating the lines to $b\to\infty$ 
one obtains the numerical estimates of the universal numbers $\S c$ which are listed in Table 1.}
\label{fig:pc_3bit}
\end{figure}

\headline{Measurement of the conditional probabilities}
Even in the parity-conserving case one can directly apply the analysis of 
section \ref{sec:blockrenorm} in order to determine the universal numbers $\S c$.
However, as an active site $i$ can be in two possible states $s_i=1,2$, 
we use the subscripts '{\tt 0}' and '{\tt a}' to denote inactive and active sites. 
For example, the renormalized density of active sites is $\P a(b)$ 
while $\S {aa}$ denotes the conditional probability of finding two consecutive 
active blocks and so on. It is then straight-forward to prove that the exact expressions
(\ref{eq:exact1}) and (\ref{eq:exact2}) of the two-bit probabilities in terms of the
critical exponents generalized to the parity-conserving case have the same
form in the new notation, i.e.
\begin{equation}
\label{eq:exactpc}
\S {0a} = \S {a0} = 1-2^{-\beta/\nu_\perp} \,,\qquad
\S {aa} = -1+2^{1-\beta/\nu_\perp} \,. 
\end{equation}

Using the estimates (\ref{eq:pc}) for the critical exponents of the PC class, we get
\begin{equation}\label{eq:pc_2bit}
\S {0a} = \S {a0}= 0.294(3)\,, \qquad \S {aa} = 0.411(2)\,.
\end{equation}
In order to verify this prediction, we simulated the PC model defined in (\ref{eq:pcrules}) with
a system size $L=10^4$. Starting from a  fully active configuration, the system evolves at the critical 
point (\ref{eq:pcpc}) for $8 \times 10^5$ Monte Carlo cycles. In certain time intervals, 
$b$ consecutive sites of the actual configuration of $L$ sites are reduced to a single 
site by applying the renormalization scheme described above. This block-renormalized configuration yields 
a string of $L/b$ sites. Again the conditional probabilities $\S c(b,t)$ are measured by scanning this string 
and averaging over many runs for block sizes $b=1,2,4,8,16,32$.

Following the same procedure as in the previous section, the left panel of 
Fig.~\ref{fig:pc_2bit} shows the two-block quantities $\S c(b,t)$ as functions of $1/t$.
The asymptotic values $\S {aa}(b)$ and $\S {a0}(b)$ obtained by extrapolating these curves
to $t \to \infty$ are then plotted against $1/b$ in the right panel of Fig.~\ref{fig:pc_2bit}.
Finally, the universal numbers $\S c= \lim_{b \to \infty} \S c(b)$ are estimated by another linear extrapolation 
of these curves. As can be seen, they are in excellent agreement with the predicted values (\ref{eq:pc_2bit}). 

We have also measured the conditional probabilities $\S c(b)$
for three adjacent blocks, for which no theoretical prediction is available.
Fig.~\ref{fig:pc_3bit} shows plot of $\S c(b)$ vs $1/b$ for different patterns ${\tt c}$,
where we distinguished only inactive and active blocks.
Note that in this case there are only five independent curves because of the
spatial reflection symmetry $\S {a00}(b,t)=\S {00a}(b,t)$ and 
$\S {aa0}(b,t)=\S {0aa}(b,t).$ Clearly, these values approach a well defined limit  
$\S c$ which is estimated by linear extrapolation as before. A summary of all
estimates is given in Table 1.
%
%
\begin{table}
\begin{center}
\begin{tabular}{|l|c|c|c|c|} \hline
$n$ & ${\tt c}$ & ${\tt c'}$ & $\S c$ (DP class) & $\S c$ (PC class)\\ \hline
2 bits: 
& \tt 01 & \tt 10 & 	0.160310(5) &  0.294(3)\\
&\tt  11 &&        	0.67938(1) & 0.411(2) \\ \hline
3 bits: 
& \tt 001 & \tt 100 	& 0.097(1) & 0.183(4)\\
& \tt 010 &		& 0.012(1) & 0.085(2)\\
& \tt 011 & \tt 110 	& 0.133(1) & 0.154(0)\\
& \tt 101 &		& 0.048(1) & 0.055(8)\\
& \tt 111 &		& 0.482(2) & 0.184(2)\\ \hline
4 bits: 
& \tt 0001 & \tt 1000 & 0.069(1) & \\
& \tt 0010 & \tt 0100 & 0.0062(5) & \\
& \tt 0011 & \tt 1100 & 0.083(1) & \\
& \tt 0101 & \tt 1010 & 0.0047(5) & \\
& \tt 0110 &&       	0.023(1) & no results\\
& \tt 0111 & \tt 1110 & 0.101(1) & \\
& \tt 1001 &&       	0.021(1) & \\
& \tt 1011 & \tt 1101 & 0.040(1) & \\
& \tt 1111 &&       	0.347(3) & \\ \hline
\end{tabular}
\end{center}
\caption{Estimates of the universal conditional probabilities $\S c$ for directed percolation and the parity-conserving universality class. ${\tt c'}$ denotes the spatially reflected  bit pattern. The 2-bit values were obtained by using the scaling relations (\ref{eq:exact1}),(\ref{eq:exact2}) and (\ref{eq:pc_2bit}), the other values by numerical simulations.}
\end{table}

\section{Summary}

In this paper we have introduced a simple renormalization scheme suitable for systems with absorbing states. For two-state systems the renormalization step involves a logical OR of the elementary degrees of freedom. The scheme can be applied in any space dimension to spatial as well as spatio-temporal configurations. 

The present work is restricted to the analysis of spatial configurations in 1+1 dimensions. It is shown that in a critical process with homogeneous initial conditions  all block probabilities $\P c(b,t)$ for non-vanishing configurations decay asymptotically as~$t^{-\alpha}$, differing only by their proportionality constants, which means that their ratios will tend to certain constants as $t\to\infty$. This allows us to define (non-universal) ratios $\S c(b) = \lim_{t\to\infty} \P c(b,t)/(1-\P{0\ldots 0}(b,t))$ which depend only on the block size. As the renormalization procedure has the tendency to suppress non-universal microscopic details, these quantities converge to \textit{universal} values $\S c$ in the limit of large block sizes $b\to\infty$. It is important to keep in mind that the two limits, $t\to\infty$ and $b\to\infty$, do not commute.

Using the examples of directed percolation and the parity-conserving class, we have demonstrated that the universal numbers are as robust as the critical exponent and therefore characterize the underlying universality class of the process like a fingerprint. The two-bit quantities turn out to depend on the critical exponents by exact scaling relations, providing a numerical test for the accuracy of the method. The three- and four-bit quantities, however, render larger sets of universal numbers which are independent of the critical exponents. By comparing the numerical estimates for a given model with the values listed in Table 1, this method provides a very sensitive test to which of the universality classes the phase transition belongs. As renormalization drives a system away from criticality, we note that it is crucial for the method to use a very accurate estimate of the critical point. 

The proposed method could shed new light onto nonequilibrium processes whose critical behavior is still debated in the literature. Examples include the pair contact process with diffusion and conserved sandpile models, which will be analyzed in a forthcoming paper.

\noindent\textbf{Acknowledgements}:\\
U. Basu would like to thank the Max Planck Institute for the Physics of Complex Systems 
in Dresden as well as the Faculty for Physics and Astronomy at the University of W\"urzburg, 
where parts of this work were done, for hospitality and financial support.


\end{document}